\documentclass[journal]{IEEEtran}

\usepackage{graphicx}
\usepackage{color}
\usepackage{amsfonts}
\usepackage{float}
\usepackage{amsmath}
\usepackage{amsthm}
\usepackage{epstopdf}
\usepackage{changepage}
\usepackage{latexsym}
\usepackage{tikz}
\usetikzlibrary{arrows}
\usetikzlibrary{automata}
\usetikzlibrary{er}
\usetikzlibrary{positioning}
\usetikzlibrary{shadows}
\usepackage{cases}
\usepackage{algorithm,algorithmic}

\newcommand*{\LargerCdot}{\raisebox{-0.25ex}{\scalebox{1.5}{$\cdot$}}}

\floatstyle{plain}
\newfloat{myalgo}{tbhp}{mya}
{\begin{myalgo}[#1]
\centering
\begin{minipage}{#2}
\begin{algorithm}[H]}%
{\end{algorithm}
\end{minipage}
\end{myalgo}}

\begin{document}

\title{Power System Dynamic State Estimation by Unscented Kalman Filter with Guaranteed \\ Positive Semidefinite State Covariance}

\author{Junjian~Qi,~\IEEEmembership{Member,~IEEE}
        and Kai~Sun,~\IEEEmembership{Senior Member,~IEEE}%
        \thanks{J. Qi and K. Sun are with Dept. of EECS, University of Tennessee, Knoxville, TN (e-mails: junjian.qi.2012@ieee.org and kaisun@utk.edu).
}
}

\maketitle

\begin{abstract}
In this paper an unscented Kalman filter with guaranteed positive semidefinite state covariance 
is proposed by calculating the nearest symmetric positive definite matrix in Frobenius norm and is applied to power system dynamic state estimation.
The proposed method is tested on NPCC 48-machine 140-bus system and the results validate its effectiveness.
\end{abstract}

\begin{IEEEkeywords}
Alternating projections, dynamic state estimation, nonlinear filters, numerical stability, positive semidefinite, state covariance, unscented Kalman filter.
\end{IEEEkeywords}

\section{Introduction}

\IEEEPARstart{P}{ower} system dynamic state estimation (PSDSE) has been implemented by extended Kalman filter (EKF) [\ref{ekf3}], [\ref{ekf4}], which maintains the elegant and efficient recursive update form of the Kalman filter, but suffers serious limitations due to linearization and Jacobian matrix calculation.

The unscented transformation (UT) [\ref{ut}] was developed to address the deficiencies of linearization
and the unscented Kalman filter (UKF) [\ref{ukf1}] was proposed based on UT and has been applied to PSDSE [\ref{pwukf1}]--[\ref{pwukf2}].
For UKF the state covariance is propagated and under some circumstances it cannot maintain
the positive semidefiniteness so that its square-root cannot be calculated, which makes UKF not numerically stable.

In this letter we propose an UKF procedure with guaranteed positive semidefinite state covariance, thus enhancing the numerical stability.
Section \ref{basis} introduces the proposed method and Section \ref{result} tests it on NPCC 48-machine system.

\section{Unscented Kalman Filter with Guaranteed Positive Semidefiniteness State Covariance} \label{basis}

A discrete-time nonlinear system can be described as
\begin{subnumcases} {}
\boldsymbol{x}_k=\boldsymbol{f}(\boldsymbol{x}_{k-1})+\boldsymbol{q}_{k-1} \\
\boldsymbol{y}_{k}=\boldsymbol{h}(\boldsymbol{x}_k)+\boldsymbol{r}_k
\end{subnumcases}
where $\boldsymbol{x}_k \in \mathbb{R}^n$ and $\boldsymbol{y}_k \in \mathbb{R}^p$ are states and measurements, the state mean and covariance are $\boldsymbol{m}$ and $\boldsymbol{P}$,
$\boldsymbol{f}$ and $\boldsymbol{h}$ are vectors of nonlinear functions, $\boldsymbol{q}_{k-1} \sim N(0,\boldsymbol{Q}_{k-1})$ and
$\boldsymbol{r}_k \sim N(0,\boldsymbol{R}_k)$ are Gaussian process noise and measurement noise.

\subsection{Unscented Transformation}

A total of $2n+1$ sigma points $\boldsymbol{X}$ are calculated as 
\begin{align}
\boldsymbol{X}^{(0)}&=\boldsymbol{m} \nonumber \\
\boldsymbol{X}^{(i)}&=\boldsymbol{m}+\eta\sqrt{\boldsymbol{P}},\quad i=1,\cdots,n \notag \\
\boldsymbol{X}^{(i)}&=\boldsymbol{m}-\eta\sqrt{\boldsymbol{P}},\quad i=n+1,\cdots,2n \nonumber
\end{align}
with weights
\begin{align}
\boldsymbol{W}_m^{(0)}&=\lambda/(n+\lambda) \nonumber \\
\boldsymbol{W}_c^{(0)}&=\lambda/(n+\lambda)+(1-\alpha^2+\beta) \notag \\
\boldsymbol{W}_m^{(i)}&=\boldsymbol{W}_c^{(i)}=1/(2(n+\lambda)), \quad i=1,\cdots,2n \nonumber
\end{align}
where $\eta=\sqrt{n+\lambda}$, $\lambda=\alpha^2(n+\kappa)-n$, and $\alpha$, $\beta$, and $\kappa$ are positive constants.

\subsection{Unscented Kalman Filter} \label{ukfp}

The initial state mean and covariance are $\boldsymbol{m}_0$ and $\boldsymbol{P}_0$.

\begin{enumerate} \renewcommand{\labelitemi}{$\bullet$}

\item Prediction
\begin{align}
&\boldsymbol{X}_{k-1}=\big [\underbrace{\boldsymbol{m}_{k-1} \cdots \boldsymbol{m}_{k-1}}_{2n+1} \big] +
\eta \big[\boldsymbol{0} \quad \sqrt{\boldsymbol{P}_{k-1}} \quad -\sqrt{\boldsymbol{P}_{k-1}}\, \big] \notag \\
&\hat{\boldsymbol{X}}_k=\boldsymbol{f}(\boldsymbol{X}_{k-1}) \qquad\quad\;\;\; \boldsymbol{m}_k^-=\sum\limits_{i=0}^{2n}\boldsymbol{W}_m^{(i)}\,\hat{\boldsymbol{X}}_{i,k} \notag \\
&\boldsymbol{P}_k^-=\sum\limits_{i=0}^{2n}\boldsymbol{W}_m^{(i)}\,(\hat{\boldsymbol{X}}_{i,k}-\boldsymbol{m}_k^-)
(\hat{\boldsymbol{X}}_{i,k}-\boldsymbol{m}_k^-)^T+\boldsymbol{Q}_{k-1} \notag
\end{align}

\item Update
\begin{align}
&\boldsymbol{X}_{k}^-=\big [\underbrace{\boldsymbol{m}_k^- \cdots \boldsymbol{m}_k^-}_{2n+1} \big] +
\eta \big[\boldsymbol{0} \quad \sqrt{\boldsymbol{P}_k^-} \quad -\sqrt{\boldsymbol{P}_k^-}\, \big] \notag \\
&\boldsymbol{\boldsymbol{Y}}_k^-=\boldsymbol{h}(\boldsymbol{X}_k^-) \qquad\qquad\;\;
\hat{\boldsymbol{y}}_k^-=\sum\limits_{i=0}^{2n}\boldsymbol{W}_m^{(i)}\boldsymbol{\boldsymbol{Y}}_{i,k}^- \notag \\
&\boldsymbol{P}_{\tilde{\boldsymbol{y}}_k\tilde{\boldsymbol{y}}_k}=\sum\limits_{i=0}^{2n}\boldsymbol{W}_c^{(i)}\big(
\boldsymbol{Y}_{i,k}^--\hat{\boldsymbol{y}}_k^-\big)\big(\boldsymbol{Y}_{i,k}^--\hat{\boldsymbol{y}}_k^-\big )^T + \boldsymbol{R}_k \notag \\
&\boldsymbol{P}_{\boldsymbol{x}_k\boldsymbol{y}_k}=\sum\limits_{i=0}^{2n}\boldsymbol{W}_c^{(i)}\big(
\boldsymbol{X}_{i,k}^--\boldsymbol{m}_k^-\big)\big(\boldsymbol{Y}_{i,k}^--\hat{\boldsymbol{y}}_k^-\big )^T \notag \\
&\boldsymbol{K}_k=\boldsymbol{P}_{\boldsymbol{x}_k \boldsymbol{y}_k} \boldsymbol{P}_{\tilde{\boldsymbol{y}}_k \tilde{\boldsymbol{y}}_k}^{-1}  \qquad \boldsymbol{m}_k=\boldsymbol{m}_k^-+\boldsymbol{K}_k\big(\boldsymbol{y}_k-\hat{\boldsymbol{y}}_k^-\big) \qquad\qquad\qquad \qquad\qquad\;\; \notag \\
&\boldsymbol{P}_k=\boldsymbol{P}_k^- - \boldsymbol{K}_k \boldsymbol{P}_{\tilde{\boldsymbol{y}}_k \tilde{\boldsymbol{y}}_k}\boldsymbol{K}_k^T \qquad\qquad\qquad\qquad\qquad\;\; \notag
\end{align}

\end{enumerate}

\subsection{Guaranteed Positive Semidefinite State Covariance}

The $\boldsymbol{P}_k^-$ in step 1 or $\boldsymbol{P}_k$ in step 2 in Section \ref{ukfp} should be positive semidefinite. If any of them is not, the nearest symmetric positive definite (nearSPD) matrix (necessarily positive semidefinite) in Frobenius norm can be obtained by the following nearSPD algorithm, which adapts the modified alternating projections method in [\ref{spd}] and then adds procedures to guarantee positive definite and symmetric. $\boldsymbol{P}_k^-$ or $\boldsymbol{P}_k$ is the input $\boldsymbol{X}_0$, which is converted to the output $\boldsymbol{X}$. Similar algorithm has been implemented as a R function ``nearPD".

  \begin{algorithm}[H]
  \renewcommand\thealgorithm{}
   \caption{nearSPD} \label{alg}
    \begin{algorithmic}
      \STATE -- \sc{Initialization}: \textnormal{Let $\Delta \boldsymbol{S}_0=\boldsymbol{0}$, iteration counter $i\leftarrow 0$.}
      \vspace{0.1cm}
      \STATE -- Alternating projections: \\
      \quad While $i<i_{max}$ and $||\boldsymbol{Y}_i-\boldsymbol{X}_i||/||\boldsymbol{X}_i||>\tau_{conv}$\\
      \qquad\qquad $\boldsymbol{Y}_i \leftarrow \boldsymbol{X}_i$ \quad $i \leftarrow i+1$ \quad $\boldsymbol{R}_i \leftarrow \boldsymbol{Y}_{i-1}-\Delta \boldsymbol{S}_{i-1} $ \\
      \qquad\qquad $[\boldsymbol{V},\boldsymbol{d}] \leftarrow eig(\boldsymbol{R}_i)$ \quad\quad\quad $\boldsymbol{p} \leftarrow\boldsymbol{d}>\tau_{eig}\max(\boldsymbol{d})$ \\
      \qquad\qquad $\boldsymbol{X}_i \leftarrow \boldsymbol{V}(:,\boldsymbol{p}) \LargerCdot \big [\underbrace{\boldsymbol{d}(\boldsymbol{p}) \cdots \boldsymbol{d}(\boldsymbol{p})}_{n} \big] \times \boldsymbol{V}(:,\boldsymbol{p})^T$  \\
      \qquad\qquad $\Delta \boldsymbol{S}_i=\boldsymbol{X}_i-\boldsymbol{R}_i$ \\
      \quad End
      \STATE -- Guaranteeing positive definite: \\
      \qquad\qquad $[\boldsymbol{V},\boldsymbol{d}] \leftarrow eig(\boldsymbol{X}_i)$ \quad\quad $Eps\leftarrow \tau_{posd}\max(\boldsymbol{d})$ \\
      \qquad\qquad $\boldsymbol{d}(\boldsymbol{d}<Eps)\leftarrow Eps $ \quad $\boldsymbol{diagX} \leftarrow diag(\boldsymbol{X})$\\
      \qquad\qquad $\boldsymbol{X} \leftarrow \boldsymbol{V}diag(\boldsymbol{d})\boldsymbol{V}^T$ \\
      \qquad\qquad $\boldsymbol{D} \leftarrow sqrt\big(\max(Eps,\boldsymbol{diagX})./diag(\boldsymbol{X})\big ) $ \\
      \qquad\qquad $\boldsymbol{X} \leftarrow diag(\boldsymbol{D}) \times \boldsymbol{X} \LargerCdot \big [\underbrace{\boldsymbol{D} \cdots \boldsymbol{D}}_{n} \big]$
      \STATE -- Guaranteeing symmetric: $\boldsymbol{X} \leftarrow (\boldsymbol{X}+\boldsymbol{X}^T)/2$
    \end{algorithmic}
  \end{algorithm}

\hspace{-0.5cm} Here ``$eig$" is the eigen decomposition, $\boldsymbol{V}$ is the matrix of eigenvectors, $\boldsymbol{d}$ is the vector of eigenvalues; ``$max$", ``$diag$", and ``$sqrt$" are Matlab functions; ``$\times$" is matrix product and ``$\LargerCdot$" is elementwise product; $||\boldsymbol{A}||$ is the Frobenius norm, the matrix norm of an $m\times n$ matrix $\boldsymbol{A}$ with entry $a_{ij}$ defined as
\begin{equation}
||\boldsymbol{A}||=\sqrt{\sum\limits_{i=1}^{m}\sum\limits_{j=1}^{n}|a_{ij}|^2}.
\end{equation}

\section{Simulation Results} \label{result}

The proposed method is implemented with Matlab and is tested on NPCC 48-machine
on a 3.4 GHz Intel(R) Core(TM) 
based desktop.
The basic UKF comes from EKF/UKF toolbox \cite{ekfukf}.
The generator and measurement model in Section \uppercase\expandafter{\romannumeral 3\relax}.C of \cite{model} is used. The NPCC data is from Power System Toolbox (PST) \cite{pst} and 27 generators have 4th-order model and the others have 2nd-order model.
The measurements are voltage phasors $E_t=e_R+je_I$ and current phasors $I_t=i_R+ji_I$ of the terminal buses of generator where PMUs are installed. The PMU sampling rate is 60 frames$/s$ [\ref{pwukf}], [\ref{model}].
Gaussian noise with zero mean and standard deviation of $10^{-2}$ is added to the measurements. 
The process noise setting is the same as \cite{model}.
The initial state mean is set to be the pre-fault states, which can be quite different from 
real states, thus making PSDSE very challenging.
For nearSPD $i_{max}=100$, $\tau_{conv}=10^{-6}$, and $\tau_{eig}=\tau_{posd}=10^{-7}$. 

\begin{figure}[H]
\centering
\includegraphics[width=2.9in]{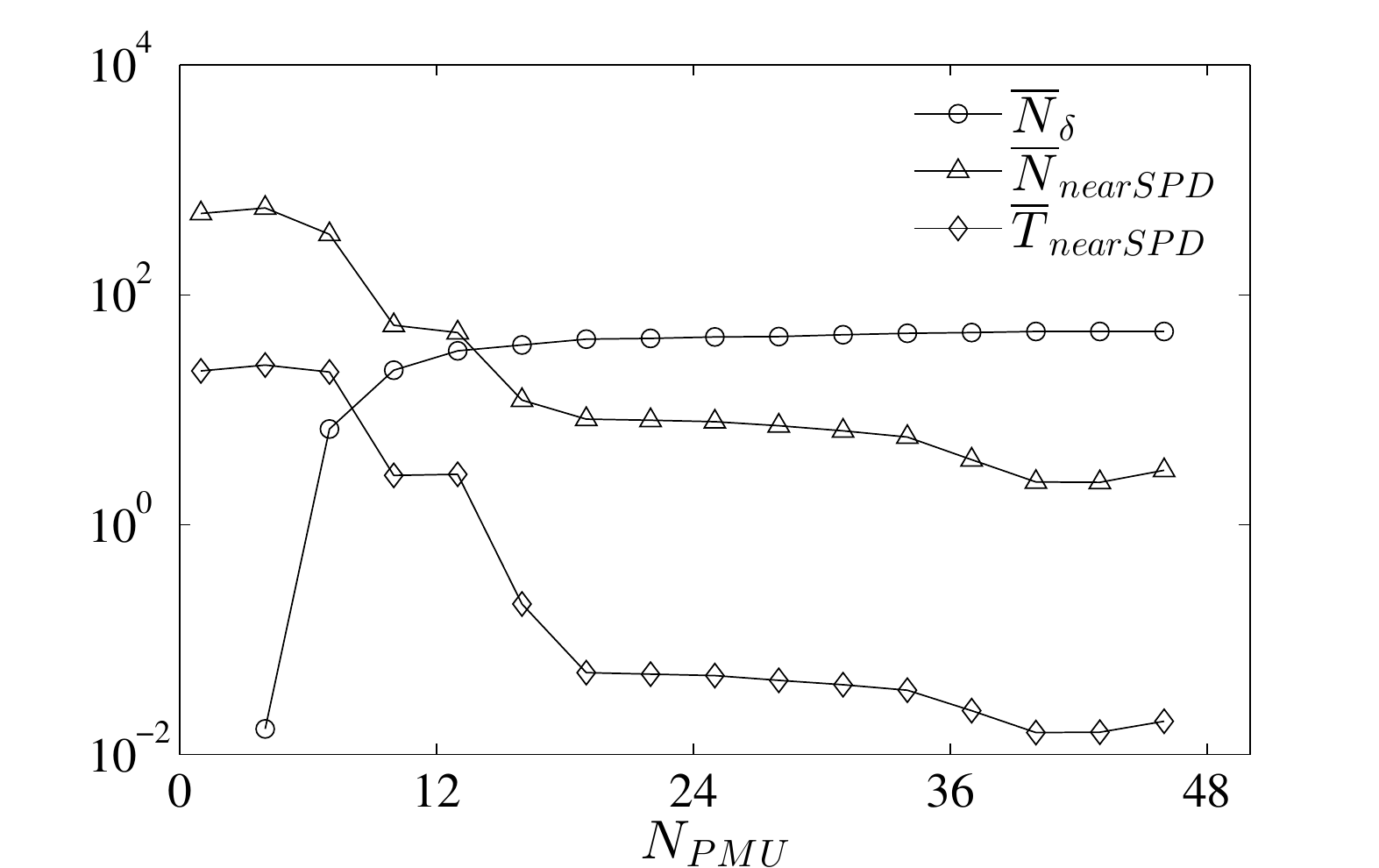}
\caption{Results for different number of PMUs.}
\label{pmu}
\end{figure}

For each number of PMUs, denoted by $N_{PMU}$, PSDSE is performed for 120 times under the optimal PMU placement in \cite{model} to estimate the system trajectory on $[0,5s]$.
For each case a fault is applied at a randomly selected location. The fault types come from PST and can be 
three-phase, line-to-ground, line-to-line to ground, line-to-line, loss of line, or loss of load at a bus. We count the number of convergent angles for which the differences between the estimated and true values in the last 0.5 second are less than 5\% of the absolute value of true values.
In Fig. \ref{pmu} we show the average ratio of convergent angles $\overline{N}_\delta$ and the average number of solving nearSPD $\overline{N}_{nearSPD}$, which respectively increases and decreases with the increase of $N_{PMU}$. When $N_{PMU}$ increases, the degree of observability of the system states also increases \cite{model} and the chance that the state covariance becomes negative definite decreases. Thus the need for solving nearSPD decreases. But for all $N_{PMU}$ there is $\overline{N}_{nearSPD}>0$, indicating that for all estimations nearSPD must be solved to make UKF work.

We also show the average time $\overline{T}_{nearSPD}$ (second) for solving nearSPD in one estimation, which gradually decreases to a very low level (less than $0.05s$) with the increase of $N_{PMU}$.  
$\overline{T}_{nearSPD}$ is a small ratio of the total estimation time: for $1\le n_{PMU}\le 10$ the ratio is about 20\%, for $10< n_{PMU}\le 13$ the ratio is about 4\%, for $n_{PMU}\ge 19$ the ratio becomes less than 0.1\%, and the smallest ratio is only 0.0193\%. Also the average time for solving nearSPD for a $150\times 150$ matrix (there are 150 states) once is only $0.0553s$.

\end{document}